# GRAVITY: getting to the event horizon of Sgr A*

F. Eisenhauer[a], G. Perrin[b], W. Brandner[c], C. Straubmeier[d], A. Richichi[f], S. Gillessen[a], J.P. Berger[e], S. Hippler[c], A. Eckart[d], M. Schöller[f], S. Rabien[a], F. Cassaing[b], R. Lenzen[c], M. Thiel[a], Y. Clénet[b], J.R. Ramos[c], S. Kellner[a], P. Fédou[b], H. Baumeister[c], R. Hofmann[a], E. Gendron[b], A. Böhm[c], H. Bartko[a], X. Haubois[b], R. Klein[c], K. Dodds-Eden[a], K. Houairi[b], F. Hormuth[c], A. Gräter[c], L. Jocou[e], V. Naranjo[c], R. Genzel[a], P. Kervella[b], T. Henning[c], N. Hamaus[a], S. Lacour[e], U. Neumann[c], M. Haug[a], F. Malbet[e], W. Laun[c], J. Kolmeder[a], T. Paumard[b], R.-R. Rohloff[c], O. Pfuhl[a], K. Perraut[e], J. Ziegleder[a], D. Rouan[b], G. Rousset[b]

[a]Max-Planck-Institut für extraterrestrische Physik, Giessenbachstraße, 85748 Garching, Germany;
[b]LESIA, Observatoire de Paris Meudon, 5, place Jules Janssen, 92195 MEUDON Cedex, France;
[c]Max-Planck-Institut für Astronomie, Königstuhl 17, 69117 Heidelberg, Germany;
[d]I. Physikalisches Institut, Universität zu Köln, Zülpicher Strasse 77, 50937 Köln, Germany;
[e]Laboratoire d'Astrophysique, Observatoire de Grenoble, BP 53, 38041 Grenoble Cédex 9, France;
[f]European Southern Observatory, Karl-Schwarzschild-Straße 2, 85748 Garching, Germany

**ABSTRACT**

We present the second-generation VLTI instrument GRAVITY, which currently is in the preliminary design phase. GRAVITY is specifically designed to observe highly relativistic motions of matter close to the event horizon of Sgr A*, the massive black hole at center of the Milky Way. We have identified the key design features needed to achieve this goal and present the resulting instrument concept. It includes an integrated optics, 4-telescope, dual feed beam combiner operated in a cryogenic vessel; near infrared wavefront sensing adaptive optics; fringe tracking on secondary sources within the field of view of the VLTI and a novel metrology concept. Simulations show that the planned design matches the scientific needs; in particular that 10µas astrometry is feasible for a source with a magnitude of K=15 like Sgr A*, given the availability of suitable phase reference sources.

**Keywords:** Galactic Center, Sgr A*, black hole, interferometry, VLTI, near-infrared

## 1. INTRODUCTION

The aim of GRAVITY is to offer to the observers an instrument that interferometrically combines near-infrared (NIR) light collected by the four unit telescopes of ESO's Very Large Telescope, using adaptive optics at the telescope level and fringe tracking at the interferometer level. This instrument can be used in an imaging mode, yielding an unprecedented resolution of ~3mas in the NIR for objects that can be as faint as $m_K = 18$ when using a fringe tracking star of $m_K = 10$. In its astrometric mode, GRAVITY will allow to measure distances between the fringe tracking star and a science object to an accuracy of 10µas, allowing one to measure directly the on-sky motions of many objects in a relatively short amount of time. At 100pc, a velocity of 10µas/yr corresponds to 5 m/s, at 1Mpc to 50 km/s. Such high precision astrometry will thus give the astronomical community a tool at hand, which makes it possible to watch how objects move in the local universe.

* ste@mpe.mpg.de, phone +49-89-300003839

## 2. KEY SCIENCE CASES

### 2.1 Stellar orbits around Sgr A*

The best case for the existence of an astrophysical black hole is the massive black hole (MBH) in the Galactic Center (GC). The combination of radio and NIR observations has proven beyond any reasonable doubt that at the dynamical center of the Milky Way a black hole of 4 million solar masses resides, coinciding with the radio source Sgr A*. The mass measurement became possible with the advent of high angular resolution techniques in the NIR. In particular, the combination of adaptive optics (AO) and large (8m-10m) telescopes made it possible to observe a multitude of stellar orbits moving in the gravitational potential of the MBH ([1], [2], [3]). Up to now the system can be described perfectly by a single point mass and Newtonian gravity. Nevertheless, deviations from these simple assumptions are expected to exist: A cluster of dark objects with stellar masses (e.g. neutron stars or stellar mass black holes) might well be present around Sgr A* ([4], [5]), yielding deviations from the single point mass hypothesis. Furthermore, the effects of general relativity will break the assumption of a Newtonian system. In order to detect such deviations, both high precision astrometry and spectroscopy of stars passing very close to the MBH can be used. Thus, this experiment naturally demands to go to the highest spatial resolution achievable in the NIR, namely interferometry.

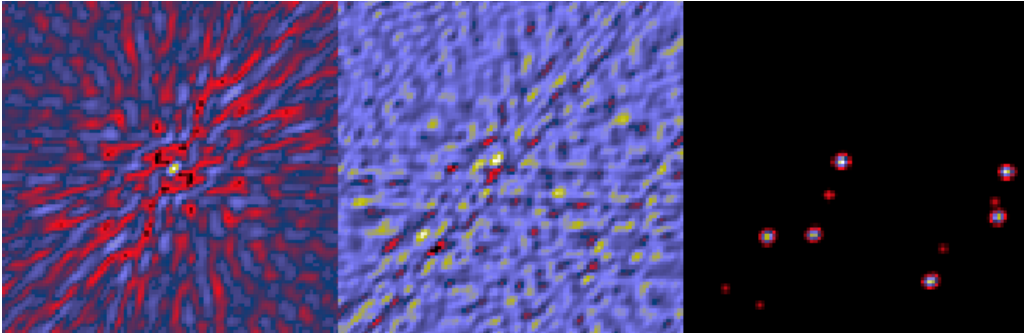

Figure 1: Simulated observation of a star field with 6 stars placed in orbit around Sgr A* in a 100mas square field. Left panel: The PSF for one night of VLTI observations. Middle panel: The reconstructed image. Right panel: The recovered image from the middle panel, using a simple CLEAN algorithm.

We have examined the feasibility of detecting the Schwarzschild precession around Sgr A*. Given the density profile and luminosity function of the Galactic Center star cluster, we estimate that a few (3 to 8) stars with $17 < m_K < 19$ at any point in time should reside in the central 100mas, thus being essentially unresolved with current NIR instrumentation, but being accessible with the VLTI. For possible applications of the upcoming two-telescope facility PRIMA in the GC see [21]. Simulating observations using all 4 UTs for 9 hours, we were able to show that it is possible to recover the assumed star fields from the simulated data. Such stars will have orbital periods of ~1year and their orbits should precess by a few degrees per revolution.

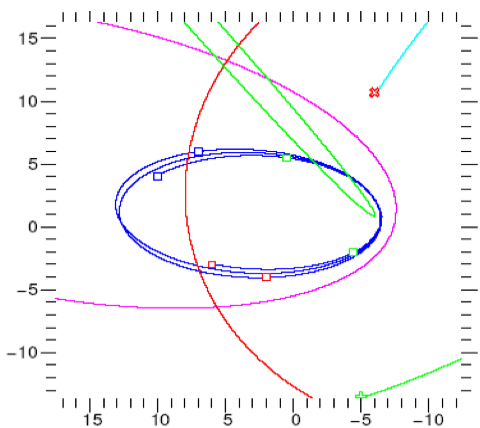

Figure 2: Simulated orbit figures for stars orbiting Sgr A* with semi major axis of ~100AU. The axes of the plot are in units of milliarcseconds. The strong precession due to the Schwarzschild metric is evident after even only two revolutions, each lasting no more than a year.

## 2.2 Flares from Sgr A*.

Genzel et al. 2003 [6] observed for the first time sporadic NIR emission from Sgr A*. Since then, many such flares occurring at a rate of 1/night and lasting each for ~2 hours have been observed. Most of the flares show a quasiperiodic substructure with a typical time scale of ~20 minutes ([7]). This can be understood in an orbiting hot spot model, where a heated gas blob close to the innermost circular orbit revolves around the MBH, yielding the light curve modulations due to the orbiting motion. Given the apparent diameter of 10μas for a MBH of 4 million solar masses at a distance of 8kpc, the motion might be detectable with NIR interferometric means.

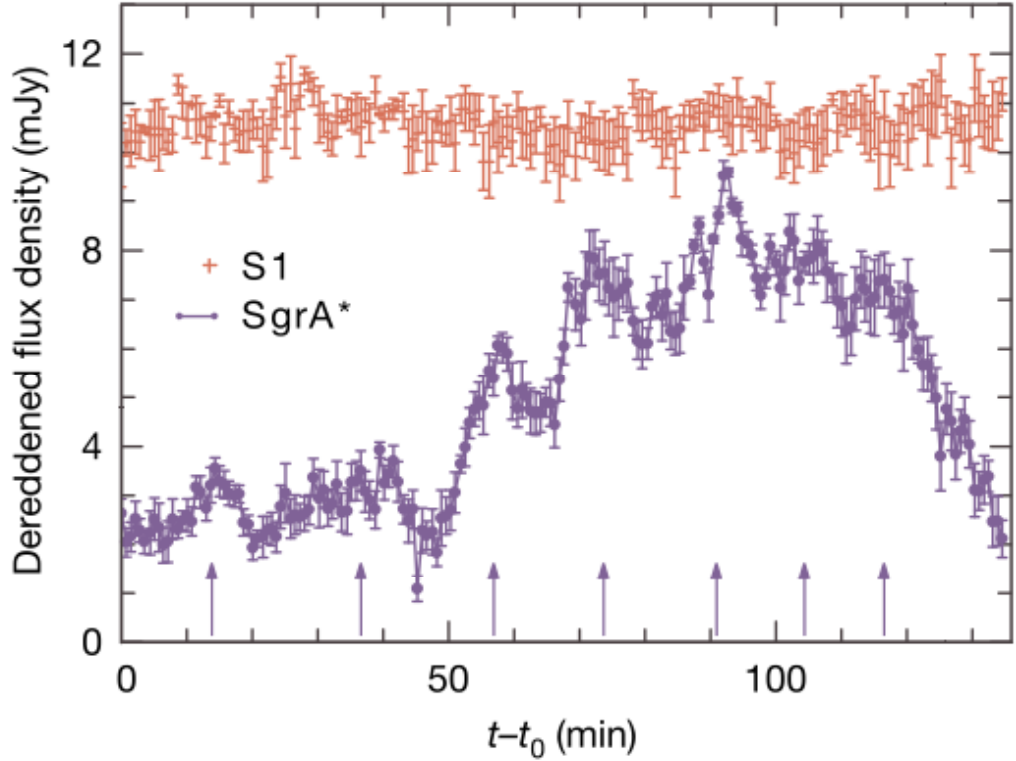

Figure 3: Light curve of a strong infrared flare, observed in K-band at the VLT on June 16, 2003. For comparison, the light curve of a star is shown. In the flare's light curve, substructure is present with a quasi-period of about 20–22 minutes. From [6].

Actually the flares will not be resolved, however the astrometric wobble of the centroid can be detected. While such an observation would already be extremely exciting, an even more rewarding goal would be to characterize the motion. The spin of the MBH and strong lensing effects will lead to characteristic deviations from a circular motion. Thus, the flares are used then as probes for the strongly curved space-time in which they move, ultimately testing the theory of gravity, general relativity (GR), in its strong field limit and for an extremely heavy mass. This perspective coined the name GRAVITY for the instrument.

We have simulated VLTI observations of flares from Sgr A*. Assuming reasonable parameters for the beam-combining instrument, we were able to show that already the observation of a single flare will allow us to detect the orbital motion. If ~10 flares can be coadded suitably, the strong relativistic effects become visible.

It should also be noted that neither the exact emission mechanism nor the exact motion are strong prerequisites. Given that flares occur close to the event horizon, their velocity should be of order speed of light, resulting in a total traveled path for a flare of one hour of ~500μas. This number is sufficiently larger than the envisioned accuracy of 10μas, showing the potential for GRAVITY to actually map the motion with high precision.

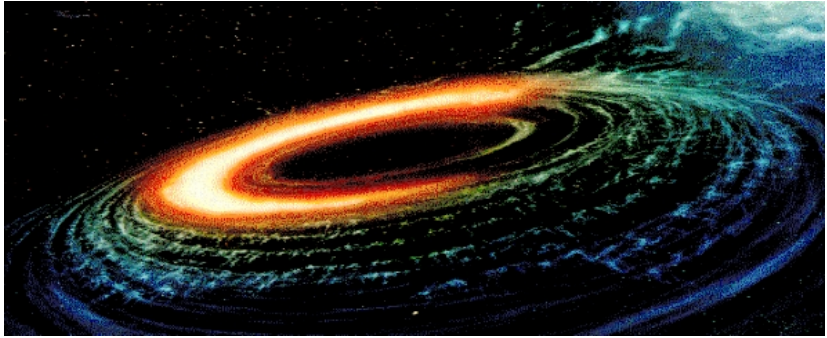

Figure 4: Illustration of the hot spot model for the flares of Sgr A*. The diameter of the orbit corresponds to ~60μas.

It is well worth to compare that route of testing GR with other experiments.

- Classical solar system tests give access to the low curvature, low mass regime of GR ([8]).
- Earth-bound gravitational wave detectors such as LIGO, VIRGO or GEO should be able to detect single supernova explosions, corresponding to the high curvature, low-mass regime of GR. The high-mass regime requires to go to lower frequencies which are not accessible from ground.
- Space-borne gravitational wave detectors (e.g. LISA) will extend the accessible frequency range to the regime in which the signal of merging MBHs is expected to occur, testing the high-curvature, high-mass regime of GR.
- A submm-VLBI array should be able to actually resolve Sgr A*, possibly showing its event horizon as a shadow ([9]).

The so far untested strong field limit of GR is tested with the last three items only. Gravitational wave detectors have a different focus, since very dynamic events will be observed. Thus the structure of space-time is tested rather indirectly. Furthermore gravitational wave detectros are quite expansive devices, and only upper limits on the emission of gravitational waves have been obtained so far experimentally. The submm-VLBI observations are unlikely to yield a dynamic picture of Sgr A* since the estimates for the exposure times needed are much longer than the orbital period, the characteristic time scale of the system. Hence, GRAVITY offers a quite promising route to severely test GR. It can be considered as a test particle approach to the structure of space-time around a MBH, which at the same time has only a very moderate price, namely that of building a 4-telescope interferometric beam combiner.

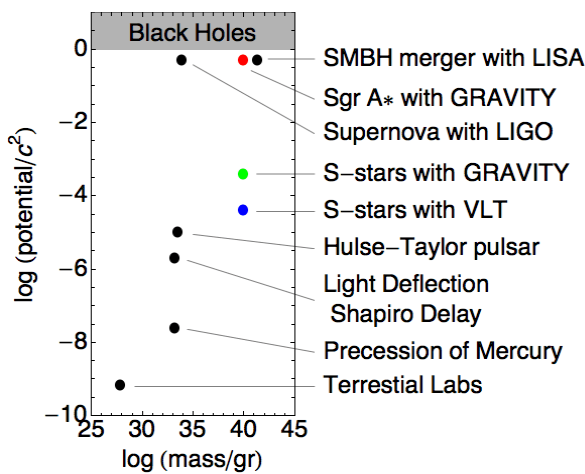

Figure 5: Dynamical tests of general relativity as a function of the mass of the gravitating body and the curvature of space-time. Adapted from [8].

## 2.3 Active galactic nuclei

The standard unified model ([10]) postulates that all active galactic nuclei (AGNs) are accreting SMBHs surrounded by a geometrically thick, dusty interstellar cloud structure (the 'torus') whose orientation relative to the observer's line of sight determines the specific phenomena observed. Historically it has proven difficult to study AGN at the spatial scales on which these components exist. For seeing limited observations, at a distance of 20Mpc, 1" corresponds to 100pc. Over the last few years, using AO for NIR observations, it has become possible to probe these objects at 100mas scales. MIR interferometry has now resolved the dusty torus for ~10 AGN and even more detailed views were possible for NGC 1068 [23], [24] and NGC 4151 [25]. GRAVITY will continue that route to smaller angular sizes. In the closest AGN – those within 20Mpc – one will be able to probe scales less than 0.5pc. And in Circinus and Cen A, the spatial resolution will be around 0.1pc, a scale which is close to what was possible in the GC before the advent of AO.

Perhaps the most exciting advances for AGN science that will become possible with GRAVITY concern the Broad Line Region (BLR). This is a compact region lying between the AGN and the inner edge of the torus, in which the large width of optical/NIR emission lines arises from the high velocities of the clouds as they orbit the central BH. Currently, the size of the BLR can only be inferred indirectly from time variability studies ('reverberation mapping', available for 35 AGN, [11]), where BLR sizes are derived from the time delay between UV continuum and emission line variations. Despite the BLR being always <0.1mas across and hence too small to resolve with VLTI direct broad-band imaging at infrared wavelengths, the excellent 10µas astrometric accuracy of GRAVITY will enable deriving a velocity gradient across it. First and foremost, this will provide a statistical estimate of the fraction of BLRs in which there is a significant component of ordered rotation. And then, for these sources, it will become possible to make an estimate of the size of the BLR and also directly determine the central BH mass.

Another important aspect is nuclear star formation. There are observational and theoretical reasons to believe that the AGN and the surrounding star formation are influencing each other. This interaction can have an impact on different scales, from fuelling the AGN (because of the effect a nuclear starburst will have on gas inflow) to the evolution of the galaxy (via feedback from the AGN). GRAVITY's spectroscopic capabilities will allow to study this on the relevant physical size scales.

## 2.4 Intermediate mass black holes

The correlation between the mass spheroidal stellar component of a galaxy and the the mass of the central MBH suggests that in dens star clusters, intermediate mass black holes could reside. This is supported by theoretical simulations that imply that in sufficiently massive and dense star clusters (e.g. [12]) the core collapses and collisionally a central object is built up. Recent searches in globular clusters suggest evidence for such IMBHs (e.g. [13], [14], [15]). However, the sphere of influence of the postulated BHs is less than a few arcseconds typically, such that only a few stars are available for this type of statistical studies. GRAVITY will dramatically change this situation in a few suitable cases. With the high angular resolution accelerations may be determined for some stars, offering a precise tool for determining the gravitational potential.

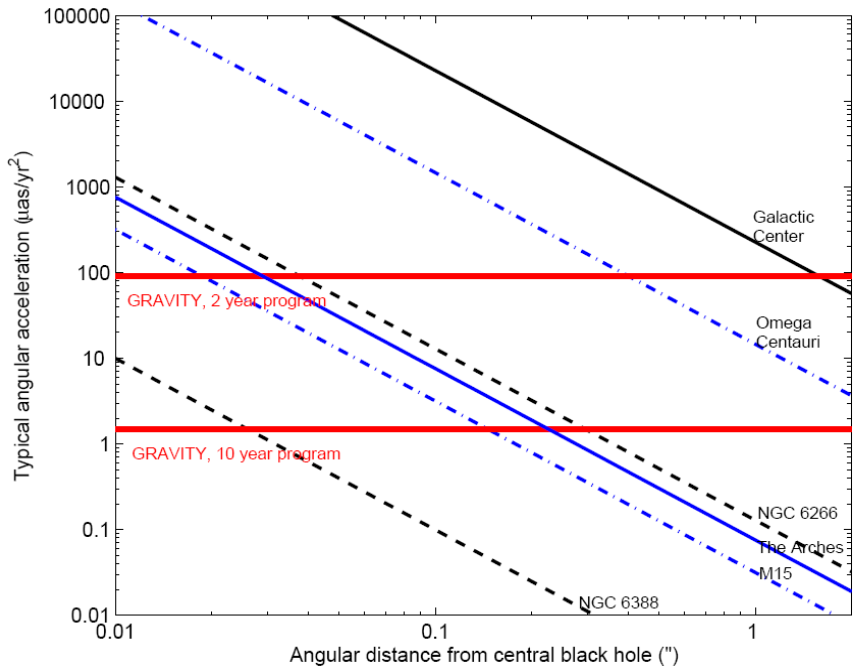

Figure 6: Angular acceleration of orbiting stars as a function of distance from the centers for several massive star clusters, assuming that these clusters contain IMBHs as estimated from recent observations.

**2.5 Young stellar objects**

GRAVITY will be ideal for investigations of YSOs, namely their circumstellar disks and outflows. Since the systems are embedded, GRAVITY's infrared wavefront-sensing capability is critical. The disk may for example show spiral structures, wakes and gaps expected as a result of the interaction between a forming planet and the disk. Such structures have indeed already been observed in a number of cases (e.g. GG Tau, Fomalhaut). GRAVITY's 4mas resolution (compared to 60mas resolution for one UT) drastically improves the contrast between a disk and its embedded planet. It should be possible to probe for young giant planets, which are almost 6 mag fainter than what is currently achievable.

The physics behind the formation of jets in YSOs, and in particular the launching mechanism, is still poorly understood. The important processes seem to take place within less than 0.5AU from the star, which at typical distances to the nearest star forming regions of 150pc translates into an angular size of less than 30mas. At 4mas resolution, GRAVITY will be able to resolve the central jet formation engine around young, nearby stars. Furthermore, at a distance of 150pc, an astrometric precision of 10µas over a time span of 1h corresponds to a transversal velocity accuracy of ≈ 60km/s. Hence, high-velocity outflows and the formation and evolution of jets from T Tauri stars with typical velocities of 150 km/s can be resolved and the time evolution of jet formation at the base of the outflow can be directly traced with GRAVITY.

# 3. TOP LEVEL REQUIREMENTS & WORKING PRINCIPLE

## 3.1 Top level requirements

Given the key science cases we have derived the following top-level requirements:

- Operation in the K-band (2.2μm), offering a well-suited atmospheric window and being the scientifically most interesting waveband for all types of obscured objects.

- Interferometric combination of the light collected by the 4 UTs of the VLT; mainly driven by the desired UV-coverage and faintness of the sources.

- Simultaneous combination of the light of two sources for each of the six baselines.

- Adaptive Optics at each telescope using a wavefront sensor operating in the NIR. A Strehl ratio of ~50% should be reached for a magnitude of $m_K$ = 6.5 assuming typical atmospheric conditions for the VLTI site, Paranal, and a distance of the guide star to the sciene field of 6". Wavefront sensing should be possible also on-source.

- Fringe Tracking, either on source or using a phase reference source within the field of view of the VLTI, down to $m_K$ = 10 assuming typical atmospheric conditions.

- Control of the relevant optical path differences at the level of 5nm, corresponding to the desired accuracy of 10μas.

- Spectral resolution between 0 and 500 and active polarization control.

## 3.2 Working principle

The working principle of GRAVITY is illustrated in figure 7. Up to four celestial sources are used in GRAVITY: The science target (SCI), the AO guide star (NGS), the fringe tracking star (FS) and an additional guide star (AGS) for compensation of tip/tilt errors occurring between the NGS and the interferometer entrance. SCI, FS and AGS have to lie within the interferometric field of view of 2" of the VLTI. The NGS can be chosen in a field of ~1' around the science field. If no AGS is available, the use of an artificial reference source for the additional tip/tilt correction is foreseen.

The light is sent from each telescope to the respective star separators, where the science field SCI and the field for the NGS are separated. Both fields are sent via the VLTI delay lines to a switchyard table located in the VLTI laboratory. At that point the path difference between the different telescopes has been compensated. The switchyard reflects the light of the NGS into the WFS modules, which in turn command the deformable mirrors of the telescopes, located just in front of the star separators.

The light from the science field enters the beam combiner instrument. In an image plane the SCI and FS are picked up by two monomode fibers. Behind this fiber coupler an acquisition camera is placed which records the image plane. During acquisition it can be used to check the position of the fibers that have to be moved to the locations of SCI and FS. During the observation, the light of the AGS (or the artificial reference source) can be seen in the acquisition camera and thus provides the possibility to measure and control the tip/tilt between the field of view and the wavefront sensor.

For each telescope the light from SCI and FS travels through the two fibers, passing a polarization control unit as well as the differential delay lines. These are needed to compensate for the differential delay occurring between SCI and FS due to their slightly deviating positions in the sky. Finally, the four SCI fibers (corresponding to the four telescopes) and the four FS fibers are fed into the science beam combiner and the fringe tracking beam combiner, both planned as integrated optics beam combiners. The output of each beam combiner consists of 24 beams, for each baseline simultaneously a full set of four phases (ABCD) is created. The beams are then imaged via a spectrometer onto the two detectors.

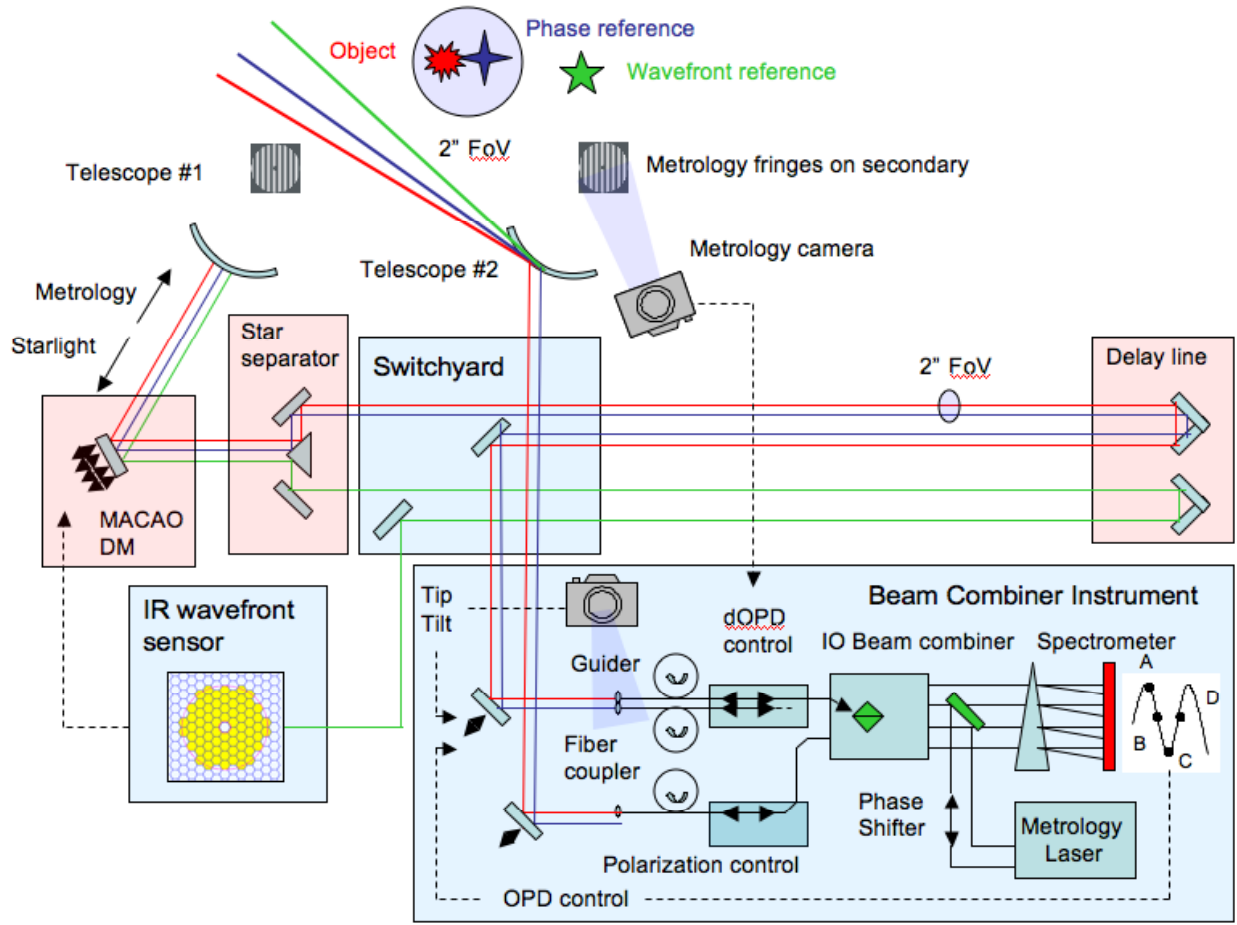

Fig. 7. GRAVITY concept. For a description see the main text.

In order to control the internal path lengths to the required accuracy of 5nm a metrology system is required. The current design foresees a novel concept for the metrology (figure 8). A NIR laser with a wavelength slightly shorter than the K-band is launched before the integrated optics beam combiners such that its light travels the whole optical path of the VLTI back to the telescopes. For each telescope the same laser light is sent once via the science channel and once via the fringe-tracking channel towards the telescope. In one of the channels a phase shifter is acting on the beam before the light enters the VLTI optics, allowing one to actively apply phase shifts between the two beams traveling towards the telescope. Due to the fixed phase relation between the two laser beams, they will create an interference pattern in each pupil plane, in particular also on M2, the secondary mirror of telescope. The fringe spacing and orientation is a function of the distance and position angle of the positions of the two fibers and thus of the celestial positions of SCI and FS. By means of the phase shifter the fringes can be sampled ('phase shifting interferometry') and the relative phase of the two beams can be recovered [19].

If the relative optical path lengths towards one telescope change (e.g. due to the operation of the differential delay line) the fringe pattern will shift by the corresponding phase. Thus, by monitoring the fringe pattern one can detect changes in the optical paths. The fringe pattern occurs on a mirror; nevertheless it will be possible to detect it in scattered light, using a commercial NIR camera watching M2 from an off-axis position in a way that M2 simply acts as a scattering screen. Our preliminary tests conducted at the VLT have shown that this manner of detection indeed is feasible and that the required accuracy can be reached.

We have simulated the interferometric performance of the instrument using realistic input data [17]. The main conclusion is that the anticipated accuracy and sensitivity can be reached. This means that the science cases as outlined above are feasible with the proposed design.

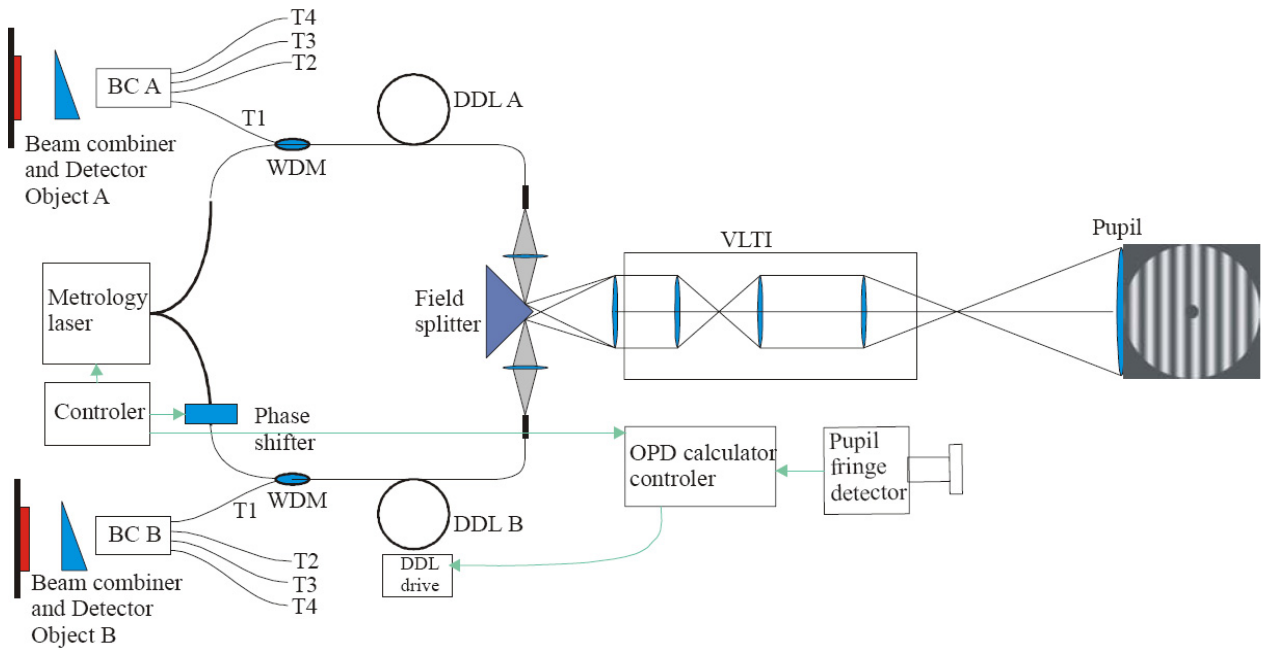

Figure 8 : Metrology concept. For a description see the main text.

## 4. PRELIMINARY DESIGN & PROTOTYPING

We briefly describe the main design features of the different subsystems of GRAVITY. The design at this stage is preliminary.

### 4.1 General Layout

The current design foresees two major components: The switchyard table and the beam combiner instrument. The switchyard table is an optical table above which the compressed VLTI beams pass and which allows one to direct light to GRAVITY's AO module and the beam combiner instrument. The latter is an evacuated cryogenic vessel, hosting the fiber optics, the actual beam combiner and the spectrometer. It has two temperature levels, one at liquid nitrogen temperature for the detectors and the other around -30°C, the value being a compromise between a low thermal background and the technical feasibility of moving stages.

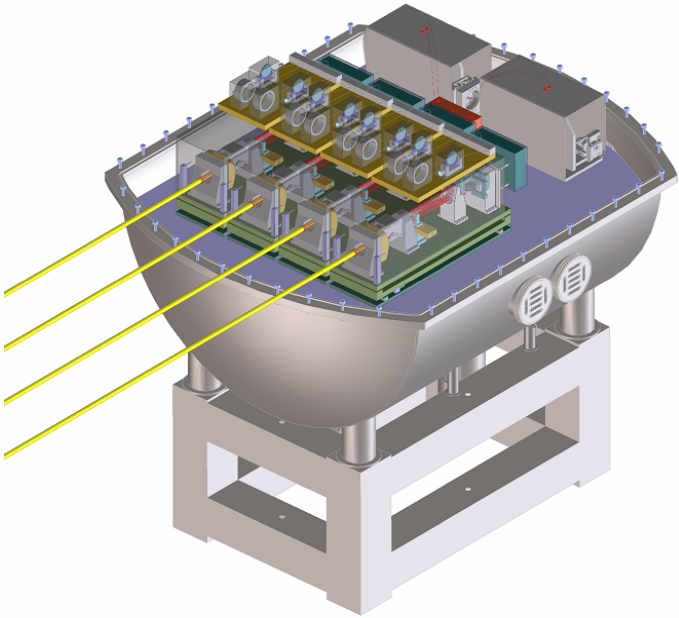

Figure 9: The GRAVITY beam combiner instrument.

### 4.2 Adaptive Optics

GRAVITY will use NIR wavefront sensing. The corresponding sensor can be located in the VLTI laboratory where actually one detector suffices to analyze the wavefronts arriving from the four telescopes. This is a cost-efficient solution since only one cryostat and one detector need to be provided. The most natural place for this 4-telescope wavefront sensor is on top of the switchyard table, which is otherwise used to pick up the beams from the VLTI lab and send them to the GRAVITY beam combiner instrument. For the baseline design we have selected Shack-Hartmann type wavefront sensors because of their robustness and uncomplicated assembly. The spatial sampling is a 9x9 square lenslet array. The deformable mirrors of GRAVITY will be the ones already installed at the Coudé labs of the individual telescopes, which are part of the MACAO adaptive optics system present at all four UTs [20].

### 4.3 Fiber optics

The 8 beams (2 objects times 4 telescopes) entering the beam combiner instrument are fed into the fiber optics units. They consist of a K-mirror for derotation, the fiber coupling units, the fibers themselves and the acquisition camera. The fiber coupling units consist of two movable fiber ends that can be moved in the image plane. The coupling of the light into the fibers is achieved by microlenses glued onto the fibers in a way that they re-image the telescope pupil onto the fiber core. The fibers serve a threefold goal: a) they are mode-cleaners, allowing only one mode of the light waves to pass; b) by fiber twisters, one can actively control the polarization characteristics of GRAVITY; and c) the fibers are wound around piezo coils, which are used to stretch them and to gain optical path length control in that way. It is planned to use standard fibers (as opposed to polarization maintaining fibers). These need to be twisted to align polarizations, a technique that has been demonstrated in FLUOR and 'OHANA.

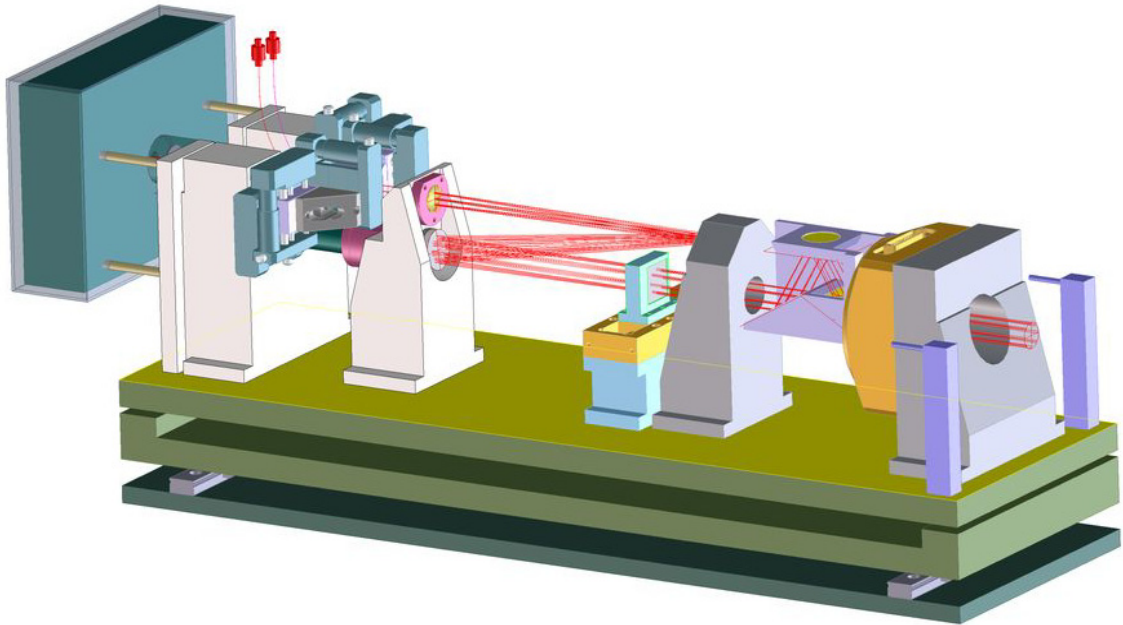

Figure 10: Preliminary design of the fiber coupling unit

## 4.4 Beam combiner

The fiber outputs are the contact points to the beam combiner, an integrated optics device. Each beam combiner has four inputs. For each of the six baselines four beam combinations are provided, each allowing for a fixed phase shift. With this scheme the beam combiner allows to sample simultaneously the phase with four points for all six baselines, yielding a total of 24 outputs [22].

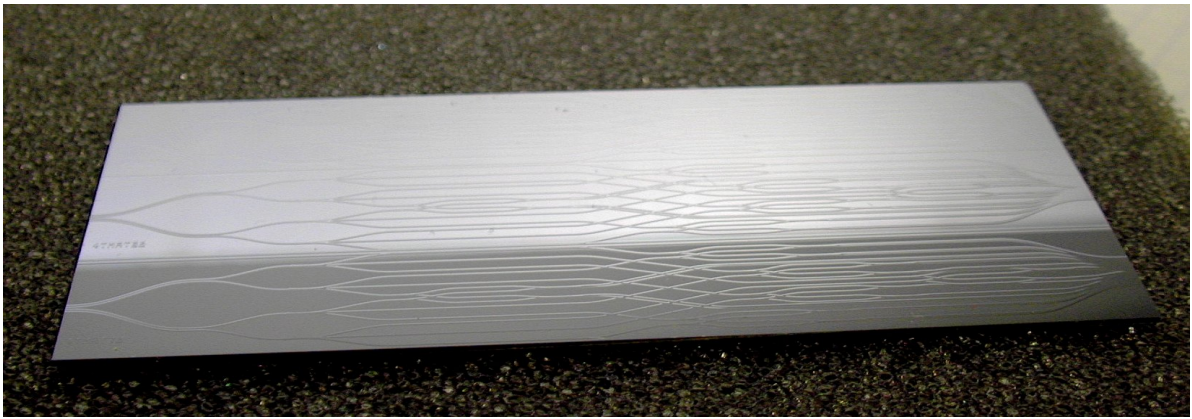

Figure 11: Integrated optics beam combiner prototype

### 4.5 Fringe Tracking

The Fringe Tracker is needed to allow for long integrations of faint signals on SCI, thus to beat down background and source photon noise as well as detector noise. The fringe tracker performance will drive the sensitivity of GRAVITY. A limiting magnitude of K=16 in 100s of integration time needs to be reached to address the science cases of the instrument. With an SNR of 10 on visibilities and an accuracy of 0.1 rad on phases (~250 nm), GRAVITY will be able to reconstruct images with details as faint as K=19 and to perform relative astrometry at 10µas accuracy.

Our study [18] shows that the requirement to fringe track with a residual rms OPD less than 300nm is met even in the stringent conditions (5 spectral channels, K=10 source, 6-baseline beam combiner, ABCD sampling). A sampling frequency of 350Hz is required. The residual OPD for atmospheric turbulence is $\lambda/12$ or 18 nm yielding a total residual (atmosphere + vibrations) of 271nm. No more than closed-loop 240nm rms opd fluctuations are allowed for VLTI vibrations in this error budget.

### 4.6 Spectrometer

The spectrometer of GRAVITY is used to disperse the 24 output beams of the beam combiner before mapping the light onto the detector. While the spectral resolution of the fringe-tracking channel is set such that the light is dispersed into five pixels, the spectral resolution of the science channel is a user-selectable parameter. Currently four different resolutions are foreseen: R=0, 32, 140, 440. An upgrade to even higher values of R seems both feasible and scientifically desirable. See also [16].

### 4.7 Metrology

The metrology system [19] will mainly be made of commercial components. The laser used for injection will operate around 1.9µm and needs a spectral stabilization since the wavelength of that laser ultimately is the ruler to which path length differences are compared. Similarly, the phase shifter for the metrology needs to be well characterized since a mismatch of assumed and actually executed phase shift would yield a systematic error on the measured phases and thus the object distances on sky.

Currently under discussion is the design of the injection optics of the laser, the main point to consider being that the relatively bright laser source should not swamp the detectors with NIR photons. Besides an intrinsically clever design the use of blocking filters seems mandatory.

The metrology receivers used to look at M2 will be commercial devices, probably upgraded with a cold-stop in order to lower the thermal noise on the detector. Also, they are equipped with a narrow line filter centered at the wavelength of the laser.

## 5. PROJECT DETAILS

GRAVITY is a joint project by currently six institutes. The German partners are the Max-Planck-Institut für extraterrestrische Physik (MPE, Garching), the Max-Planck-Institut für Astronomie (MPIA, Heidelberg) and the University of Cologne. The French team consists of the Observatoire de Paris / LESIA, ONERA and the Laboratoire d'AstrOphysique de Grenoble (LAOG). The PI-ship and project management are located at MPE that also is responsible for the beam combiner instrument cryostat, the fiber optics, the acquisition camera and the metrology. LAOG is in charge of the integrated optics beam combiner. The AO module and the switchyard are done jointly by MPIA and Observatoire de Paris / LESIA. ONERA and LESIA have formed a partnership called PHASE and are in charge of the fringe tracker and the beam combiner integration (optical functions). The spectrometer is the work package of the University of Cologne.

The project successfully concluded a feasibility study in mid 2007. Late 2007 the ESO Scientific and Technical Committee (STC) recommended to ESO to start the project. Currently, the partners work on the preliminary design and at the same time contracting with ESO is underway. Preliminary design review is foreseen for mid 2009. Shipping to Paranal of GRAVITY is planned for 2012, and science operation could start in 2013.

# REFERENCES

[1] Schödel, R. et al., Nature, Volume 419, Issue 6908, pp. 694-696 (2002).
[2] Ghez, A.M.; Salim, S.; Hornstein, S.D.; Tanner, A.; Lu, J.R.; Morris, M.; Becklin, E.E.; Duchêne, G., The Astrophysical Journal, Volume 620, Issue 2, pp. 744-757 (2005).
[3] Eisenhauer F. et al., The Astrophysical Journal, Volume 628, Issue 1, pp. 246-259 (2005).
[4] Morris, M., The Astrophysical Journal, Volume 408, Issue 2, pp. 496-506 (1993).
[5] Muno, M.P.; Pfahl, E.; Baganoff, F.K.; Brandt, W.N.; Ghez, A.; Lu, J.; Morris, M.R., The Astrophysical Journal, Volume 622, Issue 2, pp. L113-L116 (2005).
[6] Genzel, R.; Schödel, R.; Ott, T.; Eckart, A.; Alexander, T.; Lacombe, F.; Rouan, D.; Aschenbach, B., Nature, Volume 425, Issue 6961, pp. 934-937 (2003)
[7] Trippe, S.; Paumard, T.; Ott, T.; Gillessen, S.; Eisenhauer, F.; Martins, F.; Genzel, R., Monthly Notices of the Royal Astronomical Society, Volume 375, Issue 3, pp. 764-772 (2007).
[8] Psaltis, D., AIP Conference Proceedings, Vol. 714, edited by P. Kaaret, F.K. Lamb, and J.H. Swank. Melville, NY: American Institute of Physics, pp. 29-35 (2004).
[9] Falcke, H.; Melia, F.; Agol, E., The Astrophysical Journal, Volume 528, Issue 1, pp. L13-L16 (2000).
[10] Antonucci, R., Annual review of astronomy and astrophysics. Vol. 31 (A94-12726 02-90), pp. 473-521 (1993).
[11] Peterson, B.M., The Astrophysical Journal, Volume 613, Issue 2, pp. 682-699 (2004).
[12] Portegies Zwart, S.F.; Baumgardt, H.; Hut, P.; Makino, J.; McMillan, S.L.W., Nature, Volume 428, Issue 6984, pp. 724-726 (2004).
[13] Gebhardt, K.; Rich, R. M.; Ho, Luis C., The Astrophysical Journal, Volume 578, Issue 1, pp. L41-L45 (2002).
[14] Baumgardt, H.; Hut, P.; Makino, J.; McMillan, S.L.W.; Portegies Zwart, S.F., The Astrophysical Journal, Volume 582, Issue 1, pp. L21-L24 (2003).
[15] McLaughlin, D.E.; Anderson, J.; Meylan, G.; Gebhardt, K.; Pryor, C.; Minniti, D.; Phinney, S.; The Astrophysical Journal Supplement Series, Volume 166, Issue 1, pp. 249-297 (2006).
[16] Straubmeier, C.; Optomechanical design of the spectrometers of GRAVITY, Proc. SPIE 7013-109 (2008)
[17] Haubois, X.; Simulation of the interferometric performances of GRAVITY, Proc. SPIE 7013-110 (2008)
[18] Houairi, K.; Fringe tracking optimization with 4 beams: application to GRAVITY, Proc. SPIE 7013-45 (2008)
[19] Rabien, S.; Fringe detection laser metrology for differential astrometric stellar interferometers, Proc. SPIE 7013-17 (2008)
[20] Hippler, S.; Near-infrared wavefront sensing for the VLT interferometer, Proc. SPIE 7015-187 (2008)
[21] Bartko, H.; Study of the science capabilities of PRIMA in the Galactic Center, Proc. SPIE 7013-168 (2008)
[22] Lacour, S.; Characterization of integrated optics components for the second generation of VLTI instruments, Proc. SPIE 7013-40 (2008)
[23] Jaffe, W. et al., Nature, Volume 429, Issue 6987, pp. 47-49 (2004).
[24] Wittkowski, M. et al., Astronomy and Astrophysics, Volume 418, pp. L39-L42 (2004)
[25] Swain, M.: Observing NGC 4151 with the Keck Interferometer, Proc. SPIE 5491, p.1 (2004)